\documentclass[sigconf,natbib=false]{acmart}

\setcopyright{none}
\makeatletter
\renewcommand\@formatdoi[1]{\ignorespaces}
\makeatother
\renewcommand{\footnotetextcopyrightpermission}[1]{}

\usepackage{amsmath}
\usepackage{amsfonts}
\usepackage{algorithm}
\usepackage{algorithmicx}
\usepackage{algpseudocode}
\usepackage{graphicx}
\usepackage{geometry}
\usepackage{textcomp}
\usepackage{color}
\usepackage{xcolor}
\usepackage{hyperref}
\usepackage{listings}
\usepackage{tikz}
\usepackage{multirow}
\usepackage{multicol}
\usepackage{makecell}
\usepackage{booktabs}
\usepackage{utfsym}
\usepackage{wrapfig}
\usepackage{float}
\usepackage{fancyhdr}
\AtBeginDocument{%
  }

\copyrightyear{2024}
\acmYear{2024}
\setcopyright{acmlicensed}
\acmConference[DAC '24]{61st ACM/IEEE Design Automation Conference}{June 23--27, 2024}{San Francisco, CA, USA}
\acmBooktitle{61st ACM/IEEE Design Automation Conference (DAC '24), June 23--27, 2024, San Francisco, CA, USA}
\acmDOI{10.1145/3649329.3655958}
\acmISBN{979-8-4007-0601-1/24/06}

\RequirePackage[
  datamodel=acmdatamodel,
  style=acmnumeric,
  ]{biblatex}

\begin{document}

\title{SEPE-SQED: Symbolic Quick Error Detection by Semantically Equivalent Program Execution
}


\author{\textsuperscript{1,2}Yufeng Li, \textsuperscript{2}Qiusong Yang\footnotemark, \textsuperscript{2}Yiwei Ci, \textsuperscript{1,2}Enyuan Tian}
\affiliation{%
  \institution{Institute of Software, Chinese Academy of Sciences, Beijing, China\textsuperscript{1}}
  \institution{University of Chinese Academy of Sciences, Beijing, China\textsuperscript{2}}
  \country{}
}
\email{crazybinary494@gmail.com, {qiusong, yiwei}@iscas.ac.cn, tianenyuan@nfs.iscas.ac.cn}






\begin{abstract}
\textit{Symbolic quick error detection} (\textit{SQED}) has greatly improved efficiency in formal chip verification. However, it has a limitation in detecting \textit{single-instruction bugs} due to its reliance on the \textit{self-consistency} property. To address this, we propose a new variant called \textit{symbolic quick error detection by semantically equivalent program execution} (\textit{SEPE-SQED}), which utilizes program synthesis techniques to find sequences with equivalent meanings to original instructions. \textit{SEPE-SQED} effectively detects \textit{single-instruction bugs} by differentiating their impact on the original instruction and its semantically equivalent program (instruction sequence). To manage the search space associated with program synthesis, we introduce the \textit{CEGIS based on the highest priority first} algorithm. The experimental results show that our proposed CEGIS approach improves the speed of generating the desired set of equivalent programs by 50\% in time compared to previous methods. Compared to \textit{SQED}, \textit{SEPE-SQED} offers a wider variety of instruction combinations and can provide a shorter trace for triggering bugs in certain scenarios.  
\end{abstract}




\maketitle
\renewcommand{\thefootnote}{\fnsymbol{footnote}}
\footnotetext[1]{Corresponding author\\This work was supported by Basic Research Projects from the Institute of Software, Chinese Academy of Sciences (Grant No. ISCAS-JCZD-202307) and the National Natural Science Foundation of China (Grant No. 62372438).}

\section{Introduction}\label{sec: Introduction}
As technology advances and the chip market expands, there is a growing need for cost-effective and efficient chip verification methods. However, ensuring the accuracy of a processor's behavior becomes more challenging due to aggressive microarchitectural optimizations and the daunting task of considering all possible instruction interleavings.


Formal verification (FV), such as model checking \cite{clarke2018model}, excels at detecting corner cases through exhaustive design analysis. It constructs a mathematical representation of the system and formally proves desired properties. Pioneering work includes symbolic model checkers based on abstract models of microarchitecture design \cite{damm1998herbrand, berezin1998combining}. However, abstract models often overlook elusive bugs that can arise in RTL description. ISA-Formal \cite{reid2016end} for ARM processors and RISCV-Formal \cite{riscv-formal} for RISC-V processors verify designs directly from RTL descriptions, but the formulation of formal properties requires significant manual effort and substantial expertise. 

To address these issues, a groundbreaking formal verification approach called symbolic quick error detection ($SQED$) \cite{singh2018logic,singh2019symbolic,lonsing2019unlocking} has been proposed. $SQED$ utilizes model checking to prove that any instruction sequence up to a certain bound produces a correct result. It leverages the concept of design \textit{self-consistency} to establish a single universal property, which declares that the outcomes produced by both original instructions and their duplicates are identical, regardless of the specific microarchitectural design details. Therefore, $SQED$ does not require any manual property formulation. The logic bugs that can induce changes in processor architectural states can be categorized as either \textit{single-instruction} or \textit{multiple-instruction bugs}. \textit{Single-instruction bugs} refer to the erroneous behavior of a processor when executing a specific individual instruction. These bugs are independent of all previously executed instructions. \textit{Multiple-instruction bugs} refer to the erroneous behavior of a processor when executing a sequence of multiple instructions consecutively. Practical examples have demonstrated that $SQED$ is capable of efficiently detecting \textit{multiple-instruction bugs} that are otherwise difficult to detect \cite{singh2018logic,singh2019symbolic}. However, it cannot detect \textit{single-instruction bugs} \cite{lonsing2019unlocking} that affect original and duplicate instructions uniformly. Therefore, the verification process of the \textit{self-consistency} property may yield false positive results when dealing with bugs that occur within a single instruction. This motivated the work on $C\text{-}S^2QED$ \cite{devarajegowda2020gap} which formulates \textit{single-instruction} semantics by in-house metamodeling techniques \cite{devarajegowda2018meta}, but requires providing timing information about instruction in the properties, making it not microarchitecture-independent.

In this paper, we present a novel variant for $SQED$, named \textit{symbolic quick error detection by semantically equivalent program execution ($SEPE\text{-}SQED$)}. $SEPE\text{-}SQED$ determines the correctness of an implementation by verifying whether the execution of the original instruction produces consistent architectural states with the execution of its semantically equivalent program. $SEPE\text{-}SQED$ is capable of addressing both \textit{single-instruction} and \textit{multiple-instruction bugs} of the processor design. In the case of \textit{single-instruction bugs}, their effect on the original instruction and its semantically equivalent program can vary, leading to a violation of consistency. On the other hand, $SEPE\text{-}SQED$ offers a richer variety of instruction combinations compared to the singular pattern of combining original and duplicate instructions in $SQED$. As a result, $SEPE\text{-}SQED$ provides greater flexibility in triggering bugs, and in some scenarios, it can lead to shorter bug traces. The \textit{component-based counterexample-guided inductive synthesis (CEGIS)} \cite{gulwani2011synthesis, buchwald2018synthesizing} is employed to search for programs that are semantically equivalent to the original instructions. To address the vast search space associated with existing program synthesis methods, we propose the \textit{CEGIS based on the highest priority first (HPF-CEGIS)} algorithm. The experimental results demonstrate the effectiveness of our approach.

The contributions of this paper are as follows:
\begin{itemize}
    \item We improve $SQED$ by incorporating program synthesis techniques, and propose $SEPE\text{-}SQED$, which can detect all types of logic bugs that can induce changes in processor architectural states.
    \item We introduce the \textit{HPF-CEGIS} algorithm, which, compared to the previous CEGIS algorithm, synthesizes the desired program with an average reduction of 50\% in time overhead.  
    \item We verified the capability of $SEPE\text{-}SQED$ to detect two types of logic bugs through mutation testing on a real open-source high-performance processor, and the experiments demonstrated that $SEPE\text{-}SQED$ can generate bug traces shorter than $SQED$ for certain \textit{multiple-instruction bugs}.
\end{itemize}



\section{Background and Related Work}\label{sec: background}
In this section, we provide an introduction to the background knowledge of $QED$ and $SQED$ used for formal verification and program synthesis techniques. Along the way, we present the related work.

\subsection{QED and SQED}
Quick Error Detection ($QED$) \cite{DavidLin2014EffectivePV} is a testing technique that automatically transforms an existing test, which consists of a sequence of instructions, into a new test using various transformations ($QED$ transformation). These transformations, such as \textit{Error Detection using Duplicated Instructions for Validation (EDDI-V)} and \textit{Proactive Load and Check ($PLC$)}, enhance coverage and reduce error detection latency. The EDDI-V transformation is particularly relevant to this paper as it involves the duplication of instructions in an existing instruction sequence using shadow registers and memory. The \textit{design under verification (DUV)} has its registers and memory space divided into two halves, each mapped to the other through a bijective mapping. The original and duplicate instructions exclusively refer to their respective parts. In the EDDI-V transformation, every original instruction is replicated as a duplicate instruction, with the register and memory locations mapped to their corresponding values. During a $QED$ test, both the original and duplicate instruction sequences are executed from a $QED\text{-}consistent$ state, where values in corresponding registers and memory locations are identical. Duplicated instructions execute in the same relative order as the originals but may be interleaved \cite{lonsing2019unlocking}. Mismatched values between original and duplicate registers or memory locations indicate the presence of a bug trace.

Symbolic quick error detection ($SQED$) \cite{singh2018logic,lonsing2019unlocking} utilizes $QED$ principles and \textit{bounded model checking (BMC)} \cite{biere1999symbolic} to detect and localize logic bugs in RTL. $SQED$ systematically explores all possible instruction sequences of increasing length in a symbolic manner. $QED$ transformations are then applied to these enumerated instruction sequences. It automates the process of property formulation, a known challenging task, by checking a universal property (i.e., a property that is design-independent) based on $QED$ testing. For the EDDI-V  transformation, this property is referred to as self-consistency, and it can be expressed as follows:
\begin{align}\label{fml: QED property}
    QED\text{-ready} \Rightarrow QED\text{-consistent}
\end{align}
The $QED$-ready flag serves as an indicator of the successful commitment of both the original and duplicated instructions. In the context of a processor core equipped with 32 general-purpose registers, a state is deemed to be $QED$-consistent when the equality $\bigwedge_{i=0}^{15}regs[i] == regs[i+16]$ (where $regs$ represents the register file) holds. It is important to note that registers 0 to 15 correspond to the original registers, while registers 16 to 31 are their respective duplicates, following a mapping scheme where register $regs[i]$ is associated with register $regs[i+16]$.

$SQED$ with EDDI-V transformation is unable to detect \textit{single-instruction bugs} that affect both the original and duplicate instructions in a uniform manner during a $QED$ test. Consequently, the original and duplicate registers always hold the same value, leading to $QED$-consistent states.

\subsection{Program Synthesis}\label{sec: background of synthesis}
The program synthesis aims to find a program that satisfies a given specification represented as a bit-vector formula in satisfiability modulo theories (SMT) \cite{de2011satisfiability}. The problem is an \textit{exists-forall} problem expressed as follows:
\begin{align}\label{fml: program synthesis formalization}
    \exists P: \forall \Vec{I}, O: \Big(P(\Vec{I}) == O \Big) \Rightarrow \phi_{spec}(\Vec{I}, O)  
\end{align}
\eqref{fml: program synthesis formalization} indicates that if the program is executed with inputs $\Vec{I}$ and produces output $O$, then the specification $\phi_{spec}$ is satisfied. If a satisfiable result of \eqref{fml: program synthesis formalization} is obtained through the SMT solver query, then a program $P$ is synthesized. 

An effective approach for solving satisfiability problems in second-order logic ($\exists\forall$) is to employ CEGIS \cite{solar2006combinatorial} to eliminate the universal quantifier. CEGIS involves two SMT solver calls: one to construct a candidate program and another to verify its validity for all possible inputs. Gulwani et al. \cite{gulwani2011synthesis} have utilized this technique for \textit{component-based loop-free program synthesis}. They introduced first-order \textit{location variables} $L$ to establish component connections. Location variables determine the parameters of components based on their linear order. Hence, the synthesis problem is equivalent to solving the following constraint:
\begin{align*}
    \begin{split}
        &\exists L: \big(\psi_{wfp}(L) \land \forall \Vec{I}, O, Q, R: \phi_{lib}(Q,R) \land \psi_{conn}(\Vec{I}, O, Q, R, L) \Rightarrow \phi_{spec}(\Vec{I}, O)\big) \\
        &where \quad Q::=\bigcup_{j=1}^{N}\Vec{I^{j}} \quad R::=\bigcup_{j=1}^{N}{O^{j}} \quad L::=\big\{l_{x}|x \in Q \bigcup R \bigcup \Vec{I} \bigcup O \big\}
    \end{split} 
\end{align*}
Here $Q$ and $R$ collectively denote the formal inputs and outputs of $N$ components. $\phi_{lib}$ encapsulates the formula of the components, $\phi_{lib}::= \bigwedge_{j}^{N}\phi_{j}(\Vec{I^{j}}, O^{j})$. $l_{x}$ denotes the location of each variable $x$. The \textit{well-formed program constraint} $\psi_{wfp}(L)$ mandates that the inputs of each component must be either the program inputs ($\Vec{I}$) or the outputs of its preceding component, while also ensuring that the outputs of each component are distinct. Furthermore, the constraint $\psi_{conn}$ guarantees that variables sharing the same location possess identical assignments.  

However, the classical CEGIS necessitates multiple instances of each component in the library, resulting in a substantial performance overhead due to the excessive number of components. To address this, Buchwald et al. \cite{buchwald2018synthesizing} proposed an iterative CEGIS algorithm, which involves using the \textit{combinations with replacement} algorithm to select subsets of components from the library and form increasingly longer multisets. Each multiset allows for the repetition of components, and each iteration involves synthesizing the specification using the small-sized multiset. Since the goal is not to find all programs that satisfy the specification, this approach can yield shorter programs that satisfy the specification while significantly reducing the synthesis time overhead. However, the excessive number of components can lead to a large number of multisets ($\big(\binom{N}{n}\big) = \binom{N+n-1}{n}$). For instance, if there are $N = 29$ components and $n = 6$ are chosen each time to form a multiset, it would result in \textit{1344904} multisets. Moreover, many of these multisets actually cannot synthesize the specification, leading to numerous invalid solver calls. Therefore, in this paper, we propose the \textit{HPF-CEGIS} algorithm (see Section \ref{sec: HPF-CEGIS}).

\section{Overview}\label{sec: overview}
\begin{figure}[tp]
\centerline{\includegraphics[width=0.50\textwidth]{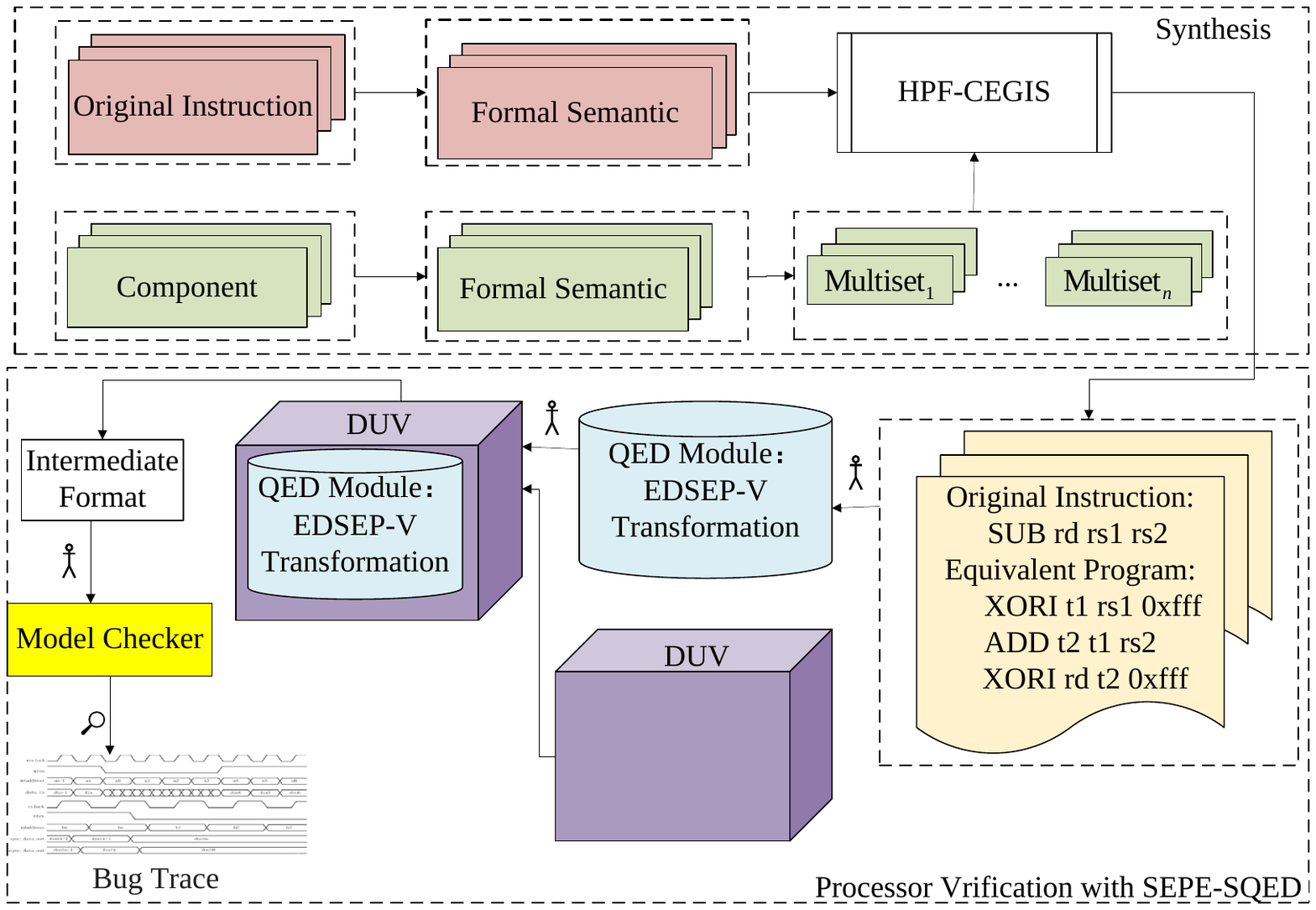}}
\caption{Workflow}
\label{fig: workflow}
\end{figure}

This section provides an overview of our approach, which involves two primary processes: synthesizing programs that are semantically equivalent to the original instructions and verifying the processor using $SEPE\text{-}SQED$. The workflow is illustrated in Figure \ref{fig: workflow}.

The upper half of Figure \ref{fig: workflow} illustrates the process of program synthesis, where the semantic models of original instructions and components defined by us (see Section \ref{sec: formal model}), are input into the program synthesizer. Subsequently, the program synthesizer invokes \textit{HPF-CEGIS} algorithm (described in Section \ref{sec: HPF-CEGIS}) to generate programs from the components that are semantically equivalent to the given original instructions. Once the instruction sequences that are semantically equivalent to the original instruction are identified, the second process involves utilizing this equivalent relation (as shown in Listing \ref{lst: SUB instruction and its semantically equivalent program}) to construct transformations. 
\lstset{
	keywordstyle=\color[RGB]{255,99,71},
	keywords={},
	frame=single,
	xleftmargin=0.5em,
	xrightmargin=0.5em,
	captionpos=b,
	breaklines=true,
	basicstyle=\footnotesize,
	caption = {\textit{SUB} instruction and its semantically equivalent program}
}
\begin{lstlisting}[mathescape,float,label={lst: SUB instruction and its semantically equivalent program}]
#Original instruction   #Semantically equivalent program
SUB rd rs1 rs2          XORI t1 rs1 0xfff
                        ADD  t2 t1 rs2 
                        XORI rd t2 0xfff   
\end{lstlisting} 

To implement $SEPE\text{-}SQED$, a special $QED$ module \cite{DavidLin2014EffectivePV} is integrated with the DUV. The module is only used for pre-silicon verification and is not added to the manufactured integrated circuit. The module takes an original instruction as input and outputs an instruction sequence that is semantically equivalent to it. Both the original and the semantically equivalent instruction sequences are fed into the DUV. Once the $QED\text{-ready}$ signal is activated, indicating the successful submission of both the original and its semantically equivalent instruction sequence, the model checker checks whether the universal property (formula ~\eqref{fml: QED property}) holds. In this paper, we refer to this transformation as \textit{Error Detection using Semantically Equivalent Program for Validation, EDSEP-V} (described in Section \ref{sec: SEPE-SQED}).

\section{Synthesis}\label{sec: synthesis}
We first present the formal semantic model of instructions (Section \ref{sec: formal model}), followed by an introduction to our \textit{HPF-CEGIS} algorithm (Section \ref{sec: HPF-CEGIS}). 

\subsection{Formal Semantic Model}\label{sec: formal model}
The synthesizer takes the semantic models of original instructions and components as inputs and utilizes these components to synthesize the original instructions that serve as the specifications (refer to formula ~\eqref{fml: program synthesis formalization}) \cite{gulwani2011synthesis, buchwald2018synthesizing}. The semantic models are represented as bit-vector formulas that precisely describe the input-output behavior of the instructions. In this paper, we use a portion of the \textit{RV32IM} \cite{waterman2014risc} instruction set as the illustrative example. 

The input and output parameters of an instruction semantic model represent register or immediate operands. They are all of the bit-vector type, but the inputs may have different bit widths. Some instructions in the library components have \textit{internal attributes} whose values are determined during synthesis. For instance, the $ADDI$ instruction has two forms as components. The first form takes both register and immediate operands as input parameters, while the second form takes only the register operand as input parameter, with the immediate operand being treated as an \textit{internal attribute}. When this component is selected, the immediate operand is assigned a specific value.  

The formal semantics of the instruction is expressed as:
\begin{align*}
    \phi_{instr}(\Vec{I}, A, O)                  
\end{align*}
where the tuple of input parameters $\Vec{I}$, the internal attribute parameter $A$ and the output parameter $O$ form the instruction's interface. For example, the semantic of \textit{ADD rd rs1 rs2} is:
\begin{align*}
    \phi_{ADD}(I_{1}, I_{2} , O) ::= (O = I_{1} + I_{2})             
\end{align*}

A library is formed by a set of specifications of components, which are expressed as:
\begin{align*}
    \big\{\langle \overrightarrow{I^{j}}, \overrightarrow{A^{j}}, O^{j}, \Phi_{j}(\overrightarrow{I^{j}}, \overrightarrow{A^{j}}, O^{j}) \rangle | j = 1, ... ,N\big\}
\end{align*} 
where all variables $\overrightarrow{I^{j}}$, $O^{j}$ are distinct, and $\Phi_{j}(\overrightarrow{I^{j}}, \overrightarrow{A^{j}}, O^{j})$ is a semantic model for one component that belongs to one of three classes
\begin{itemize}
    \item \textit{Native Instruction Class (NIC):} The semantics of the component are equivalent to the chosen instruction. For example, for the $ADD$ instruction:
    \begin{align*}
        \Phi(I_{1}, I_{2}, O) ::= \phi_{ADD}(I_{1}, I_{2}, O)
    \end{align*}
    \item \textit{Derived Instruction Class (DIC):} Derived versions of instructions can be constructed as components, where the immediate operands are treated as internal attributes rather than being taken as inputs. For example, an \textit{ADDI} instruction with a specific immediate operand can be derived as follows:
    \begin{align*}
        \Phi(I_{1}, A, O)::= \phi_{ADDI}(I_{1}, A, O)::= \Big(O = I_{1} + sext(A)\Big)  
    \end{align*}
    where $A$ is a specific 12-bit immediate operand and $sext(A)$ denotes sign-extension of $A$ to 32bits. 
    \item \textit{Composite Instruction Class (CIC):} To extend the coverage of $SEPE\text{-}SQED$ to include instructions that are difficult to synthesize under bit-vector theory, such as multiplication of two 32-bit variables, which is hard for SMT solvers, we construct the \textit{CIC} in this paper. \textit{CIC} is designed to represent the semantics of a specific instruction sequence: 
    \begin{align*}
        \Phi(\Vec{I}, \Vec{A}, O)&::=\phi_{1}(\overrightarrow{I^{1}}, A^{1}, O^{1}) \prec ... \prec \phi_{N}(\overrightarrow{I^{N}}, A^{N}, O^{N})
    \end{align*}
    where $\prec$ denotes the ordered sequence between instructions, each variable in $\overrightarrow{I^{j}}$ is either an input variable from $\Vec{I}$, or a temporary output $O^{k}$ such that $k<j$. The output of the last instruction serves as the output of the entire sequence of instructions, i.e., $O=O^{N}$. In this way, we can relax the conditions for solving. For example, to include multiplication instructions, we can allow operations that involve multiplying a 32-bit variable with a 32-bit constant:
    \begin{align*}
        \begin{split}
            &\Phi(I_{1}, A, O) ::= \phi_{ADDI}(A, O^{1}) \prec \phi_{MUL}(I_{1}, O^{1}, O) \quad \Leftrightarrow\\
            &\Phi(I_{1}, A, O) ::= \Big(O = I_{1} \times \big(0 + sext(A)\big)\Big) 
        \end{split}
    \end{align*}
\end{itemize}

To ensure that input parameters of components with different bit widths are restricted to sources of the same width, we refer to Buchwald et al.'s $\psi_{wfp}(L)$ \cite{buchwald2018synthesizing}. In addition, we introduce an \textit{input constraint} to eliminate cases where the synthesized program is identical to the original instruction $g$. The constraint can be defined as follows: 
\begin{align*}
    \begin{split}
        \Big(Name \big(\phi_{g}(\Vec{I}, O) \big) == Name \big(\Phi_{j}(\overrightarrow{I^{j}}, O^{j}) \big)\Big) \Rightarrow L(\Vec{I}) \ne L(\overrightarrow{I^{j}})
    \end{split}
\end{align*}
This constraint ensures that the synthesized program is not identical to itself, as self-equivalence would degrade into $SQED$.
 
\subsection{CEGIS Based on the Highest Priority First}\label{sec: HPF-CEGIS}
For the classical CEGIS \cite{gulwani2011synthesis}, the number of components can cause considerable performance issues. The iterative CEGIS algorithm \cite{buchwald2018synthesizing} can produce a large number of multisets through \textit{combinations with replacement} algorithm (refer to Section \ref{sec: background of synthesis}), rendering it impractical to exhaustively enumerate them within a reasonable time frame. Therefore, we propose \textit{HPF-CEGIS} (Algorithm \ref{alg: HPF-CEGIS}). 

In \textit{HPF-CEGIS}, each component $j$ is assigned a priority determined by choice weight $c_{j}$ and exclusion weight $e_{j}$. A higher $c_{j}$ indicates a higher priority, while a higher $e_{j}$ value indicates a lower priority. Initially, the weights of all components are recorded in a global dictionary (line 2). Selecting the multiset with the \textit{highest priority} (line 9, 10) before synthesis based on two factors: 
\begin{itemize}
    \item If a multiset contains some components with the same name as the original instruction, its priority is reduced to minimize the overlap between the data paths covered by the original instruction and its semantically equivalent program. For instance, we prefer using \textit{\{SUB t1 rs1 rs1, SUB t2 t1 rs2, SUB rd rs1 t2\}} instead of \textit{\{SRAI t1 rs1 0x0, ADD t2 rs2 t1, SRAI rd t2 0x0\}} to represent \textit{ADD rd rs1 rs2}.
    \item If a multiset can synthesize the original instruction, the priorities of its components are increased due to their significant semantic similarity to the original instruction. Conversely, the priorities of its components are decreased.
\end{itemize}
The calculation of the priority for a multiset with $n$ components is as follows:
\begin{align*}
    \begin{split}
        priority = \frac{\Sigma_{j=1}^{n}(c_{j}-\alpha \times \chi_{j})}{\Sigma_{j=1}^{n}e_{j}}  \quad \chi_{j} = \left \{\begin{aligned} 1& \quad Name(j) == Name(g) \\ 0& \quad Name(j) \ne Name(g) \end{aligned} \right.\\
    \end{split}
\end{align*}
Here $\chi_{j}$ is a characteristic function indicating whether the type of the component $j$ matches the original instruction $g$, and $\alpha$ is the influencing factor. 

If the synthesis fails (CEGIS returns $None$, line 12), the priorities of all components in the current multiset are reduced by increasing the values of their exclusion weights (line 13). Otherwise, the priorities of that multiset's components are enhanced by increasing the values of their choice weights (line 16). The iteration stops once the number of synthesized programs reaches a predefined threshold (line 19). 
\begin{algorithm}[tp]
\footnotesize
\caption{HPF-CEGIS}
\label{alg: HPF-CEGIS}
\begin{algorithmic}[1]
    \renewcommand{\algorithmicrequire}{\textbf{Input:}}
    \renewcommand{\algorithmicensure}{\textbf{Output:}}
    \renewcommand{\algorithmicfor}{\textbf{for}}
    \newcommand{\algorithmicendfor}{\textbf{end for}}
    \renewcommand{\algorithmicwhile}{\textbf{while}}
    \newcommand{\algorithmicendwhile}{\textbf{end while}}
    \renewcommand{\algorithmicif}{\textbf{if}}
    \newcommand{\algorithmicendif}{\textbf{end if}}

    \State \textbf{procedure} PRIORITYITERATION($G$: \{Original instruction\}, $B$: \{Component\})
        \State \hspace{1em} $PRIORITYDICT \gets$ \big\{$comp_{1}$:[$c_{1},e_{1}$], ... ,$comp_{N}$:[$c_{N},e_{N}$]\big\} \Comment{Initializing the weights of the components}
        \State \hspace{1em} $R \gets \emptyset$    
        \State \hspace{1em} \algorithmicfor \hspace{0.3em} \textbf{each} $g \in G$ \textbf{do}
                \State \hspace{3em} $MULTISETS \gets$ COMBINATIONSWITHREPLACEMENT($B, n$) \Comment{The combinations with replacement algorithm}
                \State \hspace{3em} $M \gets \emptyset$
                \State \hspace{3em} stop $\gets$ \textbf{False}
                \State \hspace{3em} \algorithmicwhile \hspace{0.3em} \textbf{not} stop \textbf{do}
                    \State \hspace{4em} SORTED($MULTISETS, PRIORITYDICT, g$) \Comment{Sorting in descending order of priority}
                    \State \hspace{4em} $S \gets MULTISETS[0]$ \Comment{The highest priority first}
                    \State \hspace{4em} $P \gets$ CEGIS($g, S$) \Comment{Generating semantically equivalent program}
                    \State \hspace{4em} \algorithmicif \hspace{0.3em} $P == None$ \textbf{then}
                        \State \hspace{5em} Increasing the exclusion weight of components in $S$
                    \State \hspace{4em} \textbf{else}
                        \State \hspace{5em} $M \gets M \cup  \{P\}$
                        \State \hspace{5em} Increasing the choice weight of components in $S$
                    \State \hspace{4em} \algorithmicendif
                    \State \hspace{4em} \algorithmicif \hspace{0.3em} LEN($M$) $ > k$ \textbf{then}
                        \State \hspace{5em} stop $\gets$ \textbf{True}
                    \State \hspace{4em} \algorithmicendif
                \State \hspace{3em} \algorithmicendwhile
                \State \hspace{3em} $R \gets R \cup \{(g, M)\}$
        \State \hspace{1em} \algorithmicendfor
    \State \textbf{end procedure}
\end{algorithmic}
\end{algorithm}

\section{Processor Verification with SEPE-SQED}\label{sec: SEPE-SQED}
The correspondences between the original instructions and their semantically equivalent programs are stored in $R$ (Algorithm \ref{alg: HPF-CEGIS} line 22). These correspondences (as shown in Listing \ref{lst: SUB instruction and its semantically equivalent program}) guide us in implementing the EDSEP-V transformation. 

Following the $QED$ consistency comparison principle, the input and output registers of the original instructions are mapped to corresponding registers in the semantically equivalent instruction sequences. Additionally, some intermediate inputs and outputs of semantically equivalent instruction sequences also require register allocation. To keep the triggering logic of the $QED\text{-}ready$ signal simple, i.e., the number of register writebacks in the original instruction sequence is equal to the number of register writebacks in the semantically equivalent instruction sequence, we divided the register file into three parts. For a processor with 32 general-purpose registers, the segmentation is as follows: 
\begin{align*}
    \begin{split}
        &O ::= \{regs[0], ... , regs[12]\}\\  
        &E ::= \{regs[13], ... , regs[25]\}  \quad \forall o \in O: \exists e \in E: o \mapsto e \\
        &T ::= \{regs[26], ... , regs[31]\}\\
    \end{split}
\end{align*}
The register allocation scheme assigns the register set $O$ to the original instructions, while the register sets $E$ and $T$ are allocated to the semantically equivalent instruction sequences. Registers in $O$ are paired one-to-one with registers in $E$, while registers in $T$ serve as intermediate inputs and outputs. To maintain the data dependencies of the original instructions, the allocation of registers in $T$ must adhere to the read-after-write principle. According to the correspondences in Listing \ref{lst: SUB instruction and its semantically equivalent program}, the transformation of \textit{SUB rd rs1 rs2} is depicted in Listing \ref{lst: EDSEP-V transformation}.  
\lstset{
	keywordstyle=\color[RGB]{255,99,71},
	keywords={},
	frame=single,
	xleftmargin=0.5em,
	xrightmargin=0.5em,
	captionpos=b,
	breaklines=true,
	basicstyle=\footnotesize,
	caption = {EDSEP-V transformation}
}
\begin{lstlisting}[mathescape,float,label={lst: EDSEP-V transformation}]
#Original instruction
SUB regs[1](rd) regs[2](rs1) regs[3](rs2)    

#Semantically equivalent instruction sequence
XORI regs[26](t1) regs[15](rs1) 0xfff
ADD regs[27](t2) regs[26](t1) regs[16](rs2)
XORI regs[14](rd) regs[27](t2) 0xfff
\end{lstlisting} 

\begin{figure}[tp]
\centerline{\includegraphics[width=0.50\textwidth]{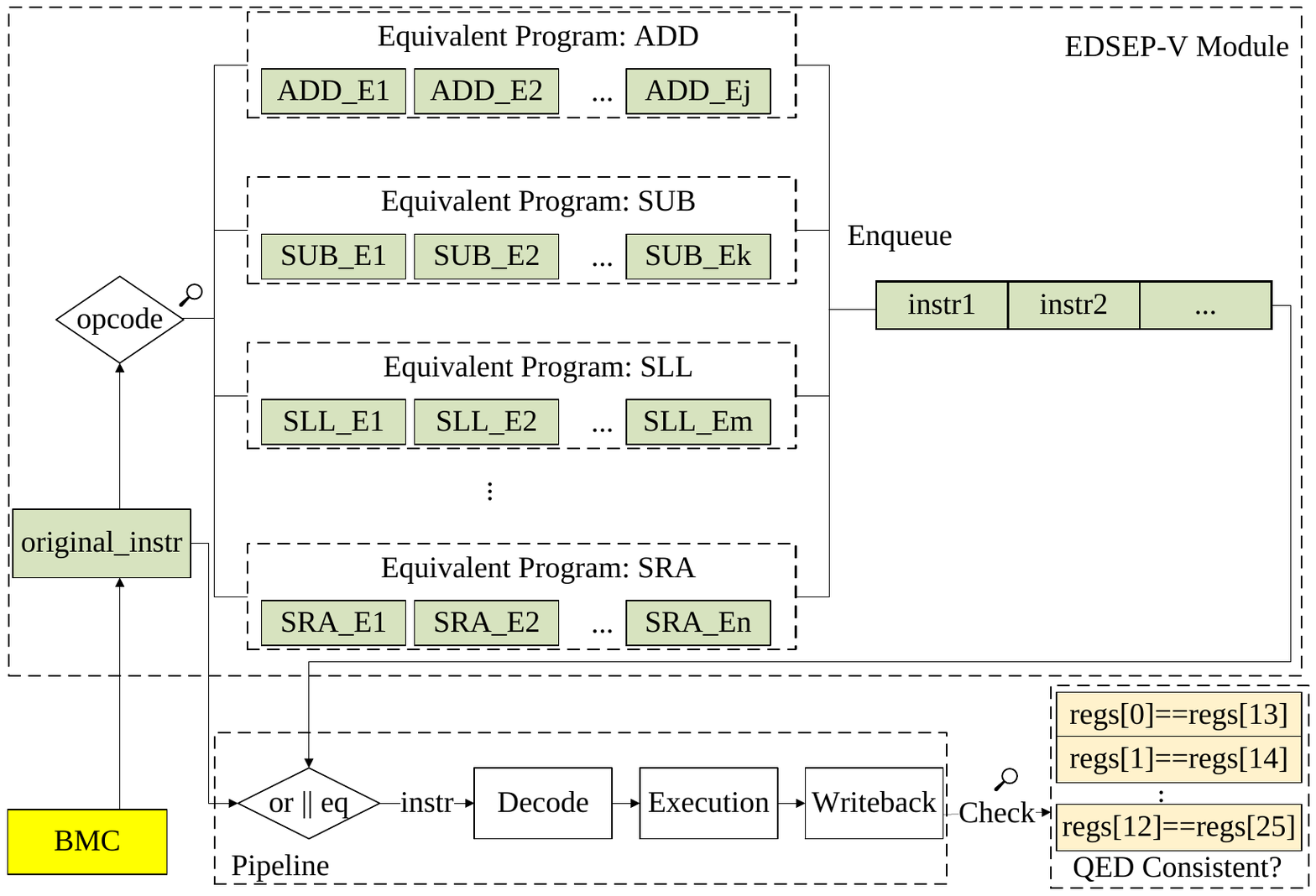}}
\caption{$SEPE$-$SQED$ verification model}
\label{fig: EDSEP-V}
\end{figure}

Figure \ref{fig: EDSEP-V} illustrates the integration of the EDSEP-V module into the DUV's RTL during verification. The solver symbolically enumerates the original instructions under the ISA for execution, while concurrently the EDSEP-V module transforms them into corresponding semantically equivalent instruction sequences. These semantically equivalent instruction sequences are stored in a queue. Based on a selection signal ($or || eq$), choose to dispatch the original instruction or semantically equivalent instruction into the pipeline. When the number of committed original instructions and their semantically equivalent counterparts is the same (determined by the number of register write-backs belonging to $O$ or $E$), the model checker checks whether the state is $QED$-consistent:
\begin{align*}
    QED\text{-ready} \Rightarrow \bigwedge_{i=0}^{12}regs[i]==regs[i+13]
\end{align*}

$SEPE\text{-}SQED$ can detect all types of logic bugs that can induce changes in processor architectural states in pre-silicon verification. On one hand, it shares the capability of $SQED$ to systematically enumerate different combinations of instructions, effectively constructing various conditions that trigger \textit{multiple-instruction bugs}. On the other hand, the original instructions and their semantically equivalent counterparts have the same functionality but different structures, thereby avoiding a single instruction bug simultaneously affecting both the original instructions and semantically equivalent sequences, leading to false positives.

\section{Evaluation}\label{sec: evaluation}
Our experimental evaluation consists of two parts. The first part involves comparing the time overhead of synthesizing the desired instruction sequences between \textit{HPF-CEGIS} and two previous CEGIS approaches. The second part tests the bug discovery capabilities of $SEPE\text{-}SQED$. The experiments were conducted on an Intel(R) Core(TM) i9-10900K CPU with 64 GB RAM running at 3.70 GHz.    

\subsection{Synthesis Algorithm}
We evaluated the time performance of three CEGIS algorithms, namely HPF-CEGIS, iterative CEGIS \cite{buchwald2018synthesizing}, and classical CEGIS \cite{gulwani2011synthesis}, in synthesizing programs with equivalent semantics. Our library consists of 29 components, including 10 NICs, 10 DICs, and 9 CICs (see Section \ref{sec: formal model}). These components collectively provide functional coverage for RV32IM instruction classes. In HPF-CEGIS, we set the initial values of each component's weight [$c_{j}, e_{j}$] and influencing factor $\alpha$ to 1 and incremented the weights by 1 with each update. For each original instruction, if 20 semantically equivalent programs consisting of at least three components have been successfully synthesized, the synthesis process will terminate early. Otherwise, all possible combinations of components will be systematically enumerated.

\begin{figure}[tp]
\centerline{\includegraphics[width=0.50\textwidth]{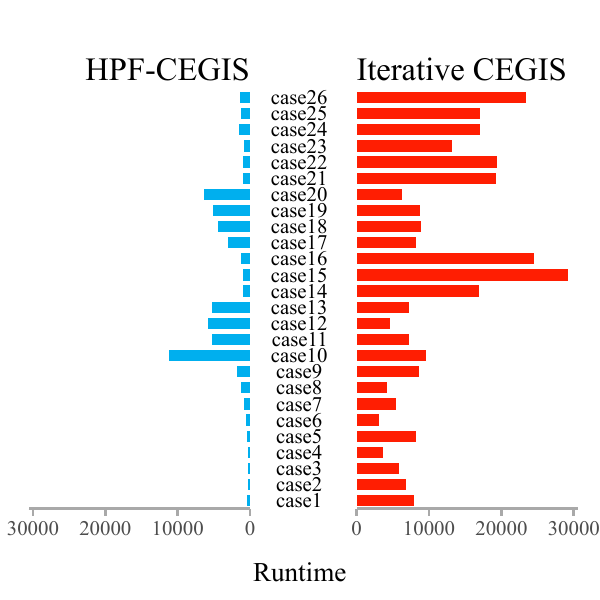}}
\caption{The time overhead of instruction synthesis}
\label{fig: synthesis algorithm}
\end{figure}

Classical CEGIS \cite{gulwani2011synthesis} failed to synthesize a single original instruction even after several weeks of experimentation with the library of 29 components. When comparing \textit{HPF-CEGIS} with iterative CEGIS, for the sake of fairness, we shuffle all multisets before synthesis to prevent the clustering of similar data types. Figure \ref{fig: synthesis algorithm} illustrates the time overhead of synthesizing different cases using two algorithms, indicating that HPF-CEGIS significantly reduces synthesis time, exhibiting an average overall reduction in synthesis time of 50\%, with synthesis time reduced by up to 90\% in certain cases.

\subsection{Real RTL Verification}
To assess the efficacy of $SEPE\text{-}SQED$ in detecting logic bugs, we conducted mutation testing on RIDECORE, an advanced superscalar and out-of-order processor core. The RTL code was converted into the BTOR2 \cite{niemetz2018btor2} intermediate format through Yosys \cite{wolf2016yosys}, and Pono \cite{DBLP:conf/cav/MannILYZBGB20} was employed as the model checking engine. 

Our approach primarily focuses on enhancing $SQED$ to extend its capability for checking \textit{single-instruction bugs}. While it is also feasible to develop specialized formal properties for \textit{single-instruction bugs} \cite{reid2016end, riscv-formal, devarajegowda2020gap}, as mentioned in the introduction (Section \ref{sec: Introduction}), these properties are non-universal and require much effort, we do not require a performance comparison with them. Table \ref{tab: Injected single-instruction bugs} demonstrates the detection results of $SEPE\text{-}SQED$ for injected \textit{single-instruction bugs} in RIDECORE.
\begin{table}[tp]
    \caption{Injected single-instruction bugs}
    \begin{center}
        \resizebox{\linewidth}{!}{
        \begin{tabular}{|l|l|c|c|}
            \hline 
            \textbf{Type}&\textbf{Function}&\textbf{SEPE-SQED}&\textbf{SQED}\\
            \hline
            ADD &Addition of two register types &3410.93s&-\\
            SUB &Subtraction of two register types &1436.46s&-\\
            XOR &Exclusive-OR &430.47s&-\\
            OR  &Bitwise OR of two register types &1765.66s&-\\
            AND &Bitwise AND of two register types &777.79s&-\\
            SLT &Set if Less Than  &3306.27s&-\\
            SLTU &Set if Less Than, Unsigned &2437.10s&-\\
            SRA &Shift Right Arithmetic &76.50s&-\\
            MULH &Multiply High &2837.13s&-\\
            XORI &Exclusive-OR Immediate &627.45s&-\\
            SLLI &Shift Left Logical Immediate &1837.11s&-\\
            SRAI &Shift Right Arithmetic Immediate &85.44s&-\\
            SW &Store Word &288.62s&-\\
            \hline
        \end{tabular}
        }
        \label{tab: Injected single-instruction bugs}
    \end{center}
\end{table}

\begin{figure}[tp]
\centerline{\includegraphics[width=0.5\textwidth]{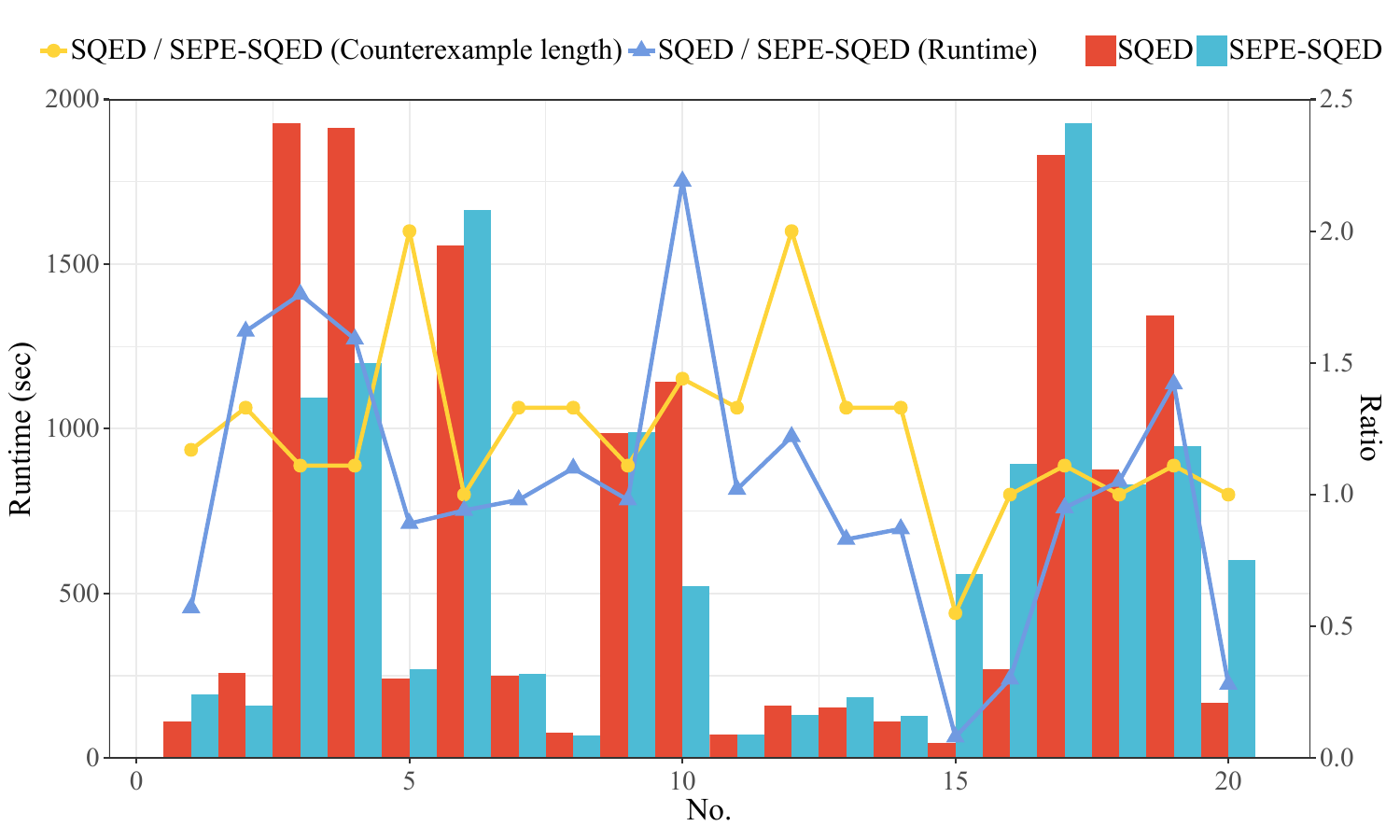}}
\caption{Detection results of multiple-instruction bugs}
\label{fig: detection_results_of_multiple-instruction_bugs}
\end{figure}

The EDSEP-V module ($SEPE\text{-}SQED$) is relatively more complex compared to the EDDI-V module ($SQED$). We also conducted tests to determine if this complexity resulted in significant overhead in detecting \textit{multiple-instruction bugs}. The x-axis of Figure \ref{fig: detection_results_of_multiple-instruction_bugs} represents the bug identifier, while the red and blue bars depict the detection time of $SQED$ and $SEPE\text{-}SQED$. The blue curve represents the detection time ratio of $SQED$ to $SEPE$-$SQED$ for the same bug, while the yellow curve represents the counterexample length ratio of $SQED$ to $SEPE$-$SQED$ for the same bug. 

Both methods are capable of detecting injected \textit{multiple-instruction bugs}. $SEPE\text{-}SQED$ not only does not incur significant time overhead, but in some cases, it exhibits shorter bug detection time and counterexample traces compared to $SQED$. We attribute this to the fact that in contrast to the single pattern of matching original and duplicated instructions in $SQED$, $SEPE\text{-}SQED$ can trigger a more diverse sequence of instructions that lead to bugs. As a result, in certain scenarios, the solver can find a shorter bug trace.

\section{Conclusion}\label{sec: conclusion}
In this paper, we propose $SEPE$-$SQED$ for processor model checking. The processor's correctness is established by comparing the consistency of its behavior with both original instructions and their semantically equivalent instruction sequences. To achieve this, program synthesis techniques are employed to discover programs that exhibit semantic equivalence to the original instructions. To improve the process of program synthesis, we present \textit{HPF-CEGIS}, an efficient CEGIS algorithm based on a highest-priority first approach. Experimental results highlight the noteworthy enhancements in program generation speed attained by \textit{HPF-CEGIS} and confirm $SEPE\text{-}SQED$'s ability to detect both \textit{single-instruction} and \textit{multiple-instruction bugs} in an open-source processor.


\begin{refcontext}[sorting = none]
\printbibliography
\end{refcontext}

\end{document}